\documentclass{article} % For LaTeX2e
\usepackage{iclr22}
\usepackage{times}

\usepackage{amsmath,amsfonts,bm}

\def\eqref#1{equation~\ref{#1}}
\def\1{\bm{1}}

\DeclareMathAlphabet{\mathsfit}{\encodingdefault}{\sfdefault}{m}{sl}
\SetMathAlphabet{\mathsfit}{bold}{\encodingdefault}{\sfdefault}{bx}{n}

\usepackage{hyperref}
\usepackage{url}
\usepackage{latexsym}
\usepackage{booktabs}
\usepackage{graphicx}
\usepackage{amsmath}
\usepackage{cleveref}
\usepackage{multirow}
\usepackage{newtxtext}
\usepackage{caption}
\usepackage{subcaption}
\usepackage{wrapfig}
\usepackage{makecell}

\newcommand{\header}[1]{\textbf{#1}\quad}
\definecolor{midnightgreen}{rgb}{0.0, 0.29, 0.33}
\definecolor{darkpink}{rgb}{0.91, 0.33, 0.5}
\definecolor{darkmagenta}{RGB}{139, 0, 139}

\newcommand{\model}{MoDIR}

\title{Zero-Shot Dense Retrieval with Momentum Adversarial Domain Invariant Representations}

\author{Ji Xin\thanks{Work partly done during the internship at Microsoft.}\\
\medskip
\\
University of Waterloo\\
\texttt{ji.xin@uwaterloo.ca}
\And
Chenyan Xiong,
Ashwin Srinivasan,
Ankita Sharma,\\
\bf Damien Jose,
Paul N. Bennett
\medskip\\
Microsoft\\
\texttt{chenyan.xiong, ashwinsr, ankita.sharma}\\
\texttt{dajose, paul.n.bennett@microsoft.com}
}

\iclrfinalcopy % Uncomment for camera-ready version, but NOT for submission.
\begin{document}

\maketitle

\begin{abstract}
Dense retrieval (DR) methods conduct text retrieval by first encoding texts in the embedding space and then matching them by nearest neighbor search.
This requires strong locality properties from the representation space, i.e, the close allocations of each small group of relevant texts, which are hard to generalize to domains without sufficient training data.
In this paper, we aim to improve the generalization ability of DR models from source training domains with rich supervision signals to target domains without any relevant labels, in the zero-shot setting.
To achieve that, we propose Momentum adversarial Domain Invariant Representation learning (\model), which introduces a momentum method in the DR training process to train a domain classifier distinguishing source versus target, and then adversarially updates the DR encoder to learn domain invariant representations.
Our experiments show that \model{} robustly outperforms its baselines on 10+ ranking datasets from the BEIR benchmark in the zero-shot setup, with more than 10\% relative gains on datasets with enough sensitivity for DR models' evaluation.
Source code of this paper will be released.

% one sentence tldr:
% We present a new method that uses mommentum adversarial learning to enforce domain invariant representations and thus improves zero-shot accuracy of dense retrieval.

% keywords: dense retrieval, zero-shot, unsupervised domain adaptation
\end{abstract}

\section{Introduction}
\label{sec:intro}

Rather than matching texts in the bag-of-words space, Dense Retrieval (DR) methods first encode texts into a dense embedding space~\citep{lee2019latent, dpr, ance} and then conduct text retrieval using efficient nearest neighbor search~\citep{sptag,scann,faiss}.
With pre-trained language models and dedicated fine-tuning techniques, the learned representation space has significantly advanced the first stage retrieval accuracy of many language systems, including web search~\citep{ance}, grounded generation~\citep{lewis2020retrieval}, open domain question answering~\citep{dpr, izacard2020leveraging}, etc.
% With pretrained language models~\citep{bert} and dedicated fine-tuning techniques~\citep{dpr, ance}, the fully learned dense representation space has significantly advanced the first stage retrieval accuracy of many language systems, including web search~\citep{ance}, open domain question answering (OpenQA)~\citep{dpr, izacard2020leveraging}, grounded language generation~\citep{lewis2020retrieval}, etc.

Purely using the learned embedding space for retrieval has raised concerns on the generalization ability, especially in scenarios without the luxury of dedicated supervision signals.
Many have observed diminishing advantages of DR models in various datasets if they are  not fine-tuned with task-specific labels, i.e., in the zero-shot setup~\citep{beir}.
However, in many scenarios outside commercial web search, zero-shot is the norm.
Obtaining training labels is difficult and sometimes infeasible, for example, in the medical domain where annotation requires strong expertise or is even prohibited because of privacy constraints.
The lack of zero-shot ability hinders the democratization of advancements in dense retrieval from data-rich domains to everywhere else. 
Many equally if not more important real-world search scenarios still rely on unsupervised exact match methods like BM25, which are developed decades ago~\citep{bm25}.

Even within the search system, generalization ability of first stage DR models is notably worse than subsequent reranking models~\citep{beir}. 
Reranking models, similar to many classification models, only require a decision boundary between relevant and irrelevant query--document pairs (q--d pairs) in the representation space.
In comparison, DR needs good local alignments in the entire space to support nearest neighbor matching, which is much harder for representation learning.

\begin{wrapfigure}{r}{0.5\textwidth}
  \begin{center}
    \vspace{-19pt}
    \includegraphics[width=0.24\textwidth]{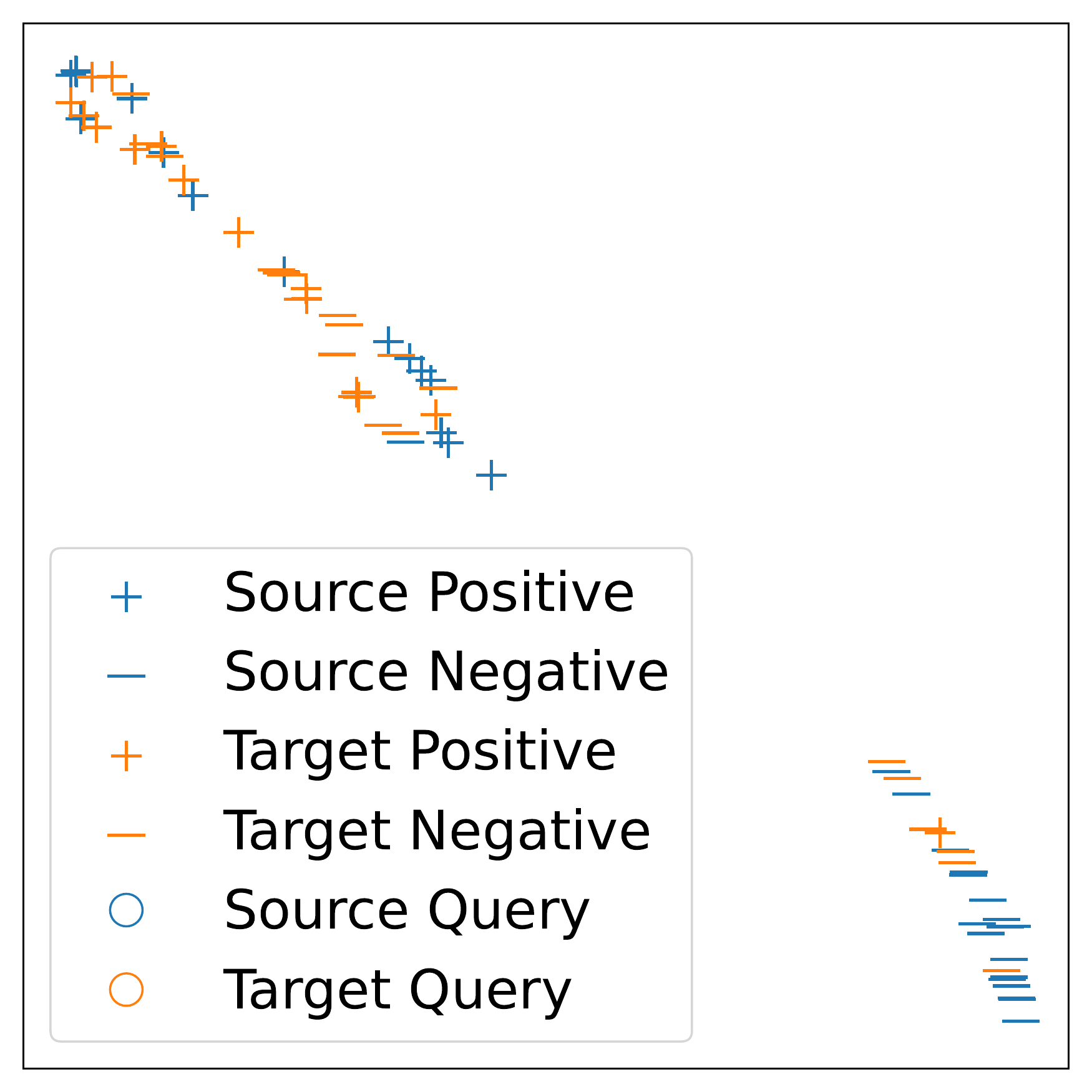}
    ~
    \includegraphics[width=0.24\textwidth]{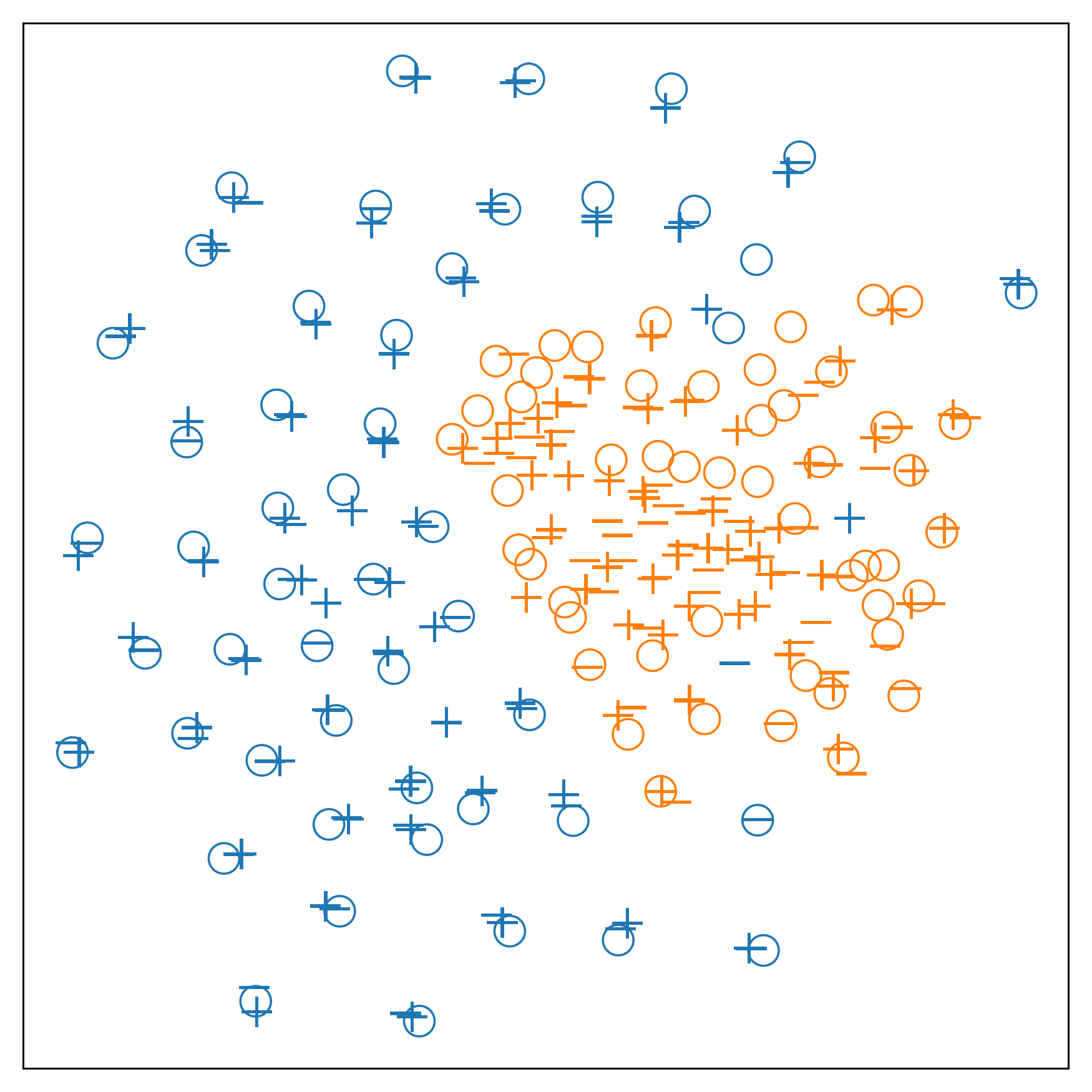}
  \end{center}
  \caption{T-SNE plots of embedding space of a BERT reranker for q--d pairs and ANCE dense retriever for queries/documents. All models are trained on web search as the source domain and applied on medical search as the target domain.
  }
  \label{fig:rerank-dr-compare}
\end{wrapfigure}

In \Cref{fig:rerank-dr-compare}, we use t-SNE~\citep{tsne} to illustrate this challenge.
We show learned representations of a BERT-based reranker~\citep{nogueira2019passage} and a BERT-based dense retriever~\citep{ance}, in zero-shot transfer from the web domain~\citep{msmarco} to medical~\citep{treccovid}.
The representation space learned for reranking yields two manifolds with a clear decision boundary; data points in the target domain naturally cluster with their corresponding classes from the source domain, leading to good generalization.
In comparison, the representation space learned for DR is more scattered.
Target domain data points are grouped separately from those of the source domain; it is nearly impossible for the learned nearest neighbor locality to generalize from source to the isolated target domain region.

% \end{figure}

In this paper, we  present \textbf{Mo}mentum Adversarial \textbf{D}omain \textbf{I}nvariant \textbf{R}epresentations learning (\model), to improve the generalization ability of zero-shot dense retrieval (ZeroDR).
We first introduce an auxiliary domain classifier that is trained to discriminate source embeddings from target ones.
Then the DR encoder is not only updated to encode queries and relevant documents together in the source domain, but also trained adversarially to confuse the domain classifier and to push for a more domain invariant embedding space.
To ensure stable and efficient adversarial learning
we propose a \textit{momentum} method that trains the domain classifier with a momentum queue of embeddings saved from previous iterations.

% \cx{overall result and momentum gain}
Our experiments evaluate the generalization ability of dense retrieval with \model{} using 15 retrieval tasks from the BEIR benchmark~\citep{beir}.
On these retrieval tasks from various domains including biomedical, finance, scientific, etc., \model{} significantly improves the zero-shot accuracy of ANCE~\citep{ance}, a recent state-of-the-art DR model trained with web search data.
Without using any target domain training labels, the improvements from \model{} are stable, robust, and also significant on tasks where evaluation labels have sufficient coverage for DR~\citep{beir}.
Our studies also verify the necessity of our momentum approach, without which the domain classifier fails to capture the domain gaps, and the adversarial training does not learn domain invariant representations, resulted in little improvement in ZeroDR.

Our further analyses reveal several interesting behaviors of \model{} and its learned embedding space.
During the adversarial training process, the target domain embeddings are gradually pushed towards the source domain and eventually absorbed as a subgroup of the source.
In the learned representation space, our manual examinations find various cases where a target domain query is located close to source queries resembling similar information needs.
This indicates that ZeroDR's generalization ability comes from the combination of information overlaps of source/target domains, and \model{}'s ability to identify the right correspondence between them.

The rest of this paper is organized as follows:
The next section presents how \model{} learns domain invariant representations for ZeroDR;
\Cref{sec:setup} and \Cref{sec:results} discuss our experimental settings and evaluation results;
We recap related works in \Cref{sec:related} and conclude in \Cref{sec:conclusion}.

\section{Training Domain Invariant Representations for Dense Retrieval}
\label{sec:method}

In this work, we aim to improve the zero-shot ability of DR in the unsupervised domain adaptation setting (UDA)~\citep{long2016unsupervised}:
Given a source domain with sufficient training signals, the goal is to transfer the DR model to a target domain, with access to its data but not any label. 
This is the common case when applying DR in real-world scenarios:\ in target domains (e.g., medical), example queries and documents are available but relevance annotations require domain expertise, while in the source domain (e.g., web search), training signals are available at large scale~\citep{ma2020zero, beir}. 

Our method, \model{}, improves ZeroDR in the UDA setup by encouraging the DR models to learn a domain invariant representation space to facilitate the generalization from source to target.
The rest of this section describes how to train a \textit{dense retrieval model} (Sec.~\ref{sec:dr}), how to use a \textit{momentum domain classifier} (Sec.~\ref{sec:discrimination}) to distinguish the two domains, and how to \textit{adversarially train}  (Sec.~\ref{sec:adv}) the DR model for more domain invariant representations.

\begin{figure}[t]
    \includegraphics[width=\columnwidth]{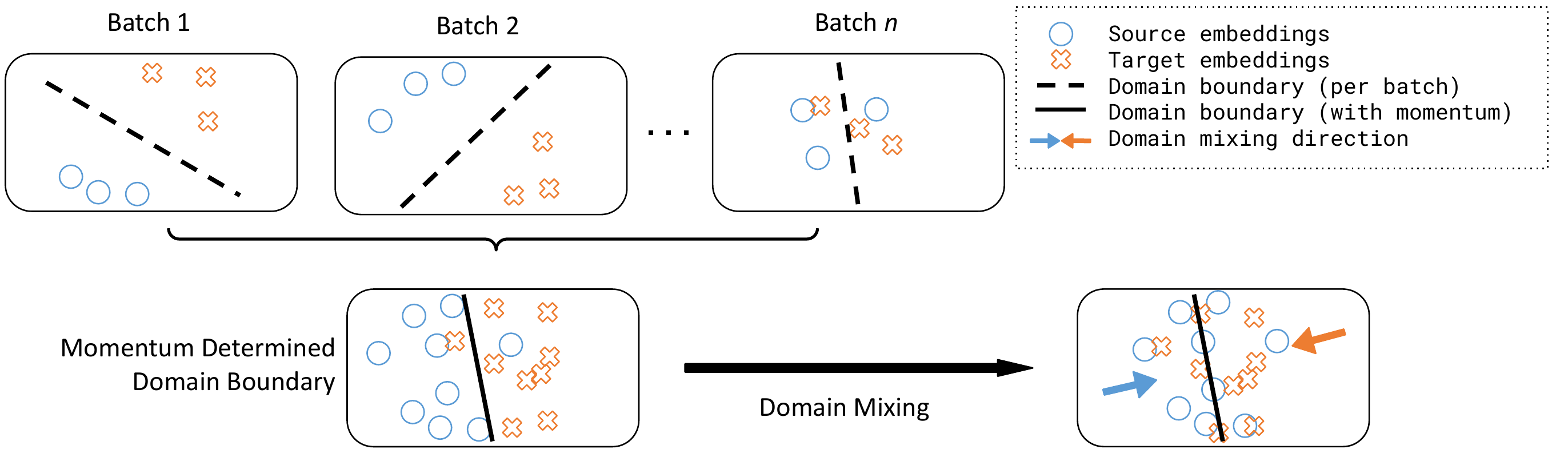}
    \caption{Momentum adversarial training provides a more accurate and robust estimation of the domain boundary in dense retrieval's embedding space.}
    \label{fig:model}
\end{figure}

\subsection{Training the Dense Retrieval Model}
\label{sec:dr}
The standard design of DR is to use a dual-encoder model~\citep{lee2019latent, dpr}, where an encoder 
$g$ takes as input a query/document and encodes it into a dense vector, and then the relevance score of a query--document pair $x=(q, d)$ is computed using a simple similarity function:
\begin{equation}
    r(x)=\mathrm{sim}(g(q;\theta_g),g(d;\theta_g)),
\end{equation}
where $\theta_g$ is the collection of parameters of $g$ and $\mathrm{sim}$ is a similarity function that supports efficient nearest neighbor search~\citep{faiss}, for example, cosine similarity or dot product.

The training of DR uses labeled q-d pairs in the source domain $x^s=(q^s,d^s)$.
% and its label $y^s$; $y^s=1$ when $q^s$ and $d^s$ are relevant and $y^s=0$ otherwise. 
With relevant q--d pair as $x^{s+}$ and irrelevant pair as $x^{s-}$,
the DR encoder $g$ is trained to minimize the \textit{ranking loss} $L_R$:
\begin{equation}
    \min_{\theta_g} \sum_{x^{s+},x^{s-}}
    L_R(r(x^{s+}), r(x^{s-})), \label{eqn:loss-ranking}
\end{equation}
where $L_R$ is a standard ranking loss function.
% , e.g., cross-entropy or negative log likelihood (NLL).
% One key component in DR training is the sampling of negative training examples, as the DR model needs to distinguish the irrelevant documents from the entire corpus, which is infeasible to enumerate in training. 
In this paper, without loss of generality, we inherit the settings of ANCE~\citep{ance} that sample negatives $x^{s-}$ using the DR model being trained. Other components are also kept the same with ANCE: $g$ is fine-tuned from RoBERTa-base~\citep{roberta} and outputs the embedding of the last layer's \textsc{[CLS]} token, $L_R$ is the Negative Log Likelihood (NLL) loss, and $\mathrm{sim}$ is the dot product.

\subsection{Estimating the Domain Boundary with Momentum Domain Classifier}
\label{sec:discrimination}
To capture the source and target domain differences and enable adversarial learning for domain invariance, 
\model{} introduces a domain classifier $f$ on top of the DR model's query/document embeddings to predict their probability of being source or target.
We simply use a linear layer on top of a data embedding $\mathbf{e}$ as the model architecture of $f$:
\begin{equation}
    f(\mathbf{e}) = \mathrm{softmax}(W_f\mathbf{e}).
\end{equation}

The linear layer is often sufficient to distinguish both domains in the high-dimensional representation space.
The challenge is mainly on the training side.
As illustrated in Figure~\ref{fig:rerank-dr-compare}, DR's  representation space focuses more on locality than forming manifolds, and therefore it is more difficult to learn the domain boundary in this case.
Learning $f$ using a large number of data points enumerated after each DR model update is costly, while updating $f$ per data batch results in an unstable estimation of domain boundary given the scattered representation space.

As shown in \Cref{fig:model}, we introduce momentum learning to balance the efficiency and robustness of the domain classifier learning.
We maintain a \textit{momentum queue} $Q$ that includes embeddings from multiple past batches as the training data for $f$.
Specifically, for each source domain training data $x^s$, 
we sample q--d pairs $x^t$ from the target domain, and add embeddings of $x^s$ and $x^t$ to $Q$. The momentum queue $Q$ at step $k$ includes embeddings from source and target for all recent $n$ batches:
\begin{equation}
    Q_k = \{ \mathbf{e}_{q^s}, \mathbf{e}_{d^s}, \mathbf{e}_{q^t}, \mathbf{e}_{d^t} |  (q^s, d^s, q^t, d^t) \in B_{k-n+1:k}\}, 
\end{equation}
where $B_{k-n+1:k}$ are the data from the past $n$ batches, with $n$ as the 
\textit{momentum step}.
We ensure the 1:1 ratio between source and target data and also 1:1 between positive and negative source data.

Note that $\mathbf{e}$ is the \textit{detached} embedding, for example, of the query $q^s$:
\begin{equation}
    \mathbf{e}_{q^s} = \Phi(g(q^s; \theta_g)),
\end{equation}
where $\Phi$ is the \textit{stop-gradient} operator, i.e., gradients of $\mathbf{e}_{q^s}$ will not be back propagated to $\theta_g$. This enables efficient momentum learning since only $W_f$ requires gradients in the process.
% and only the linear layer $W_f$ is updated. 

At each iteration, the domain classifier is updated by minimizing the following discrimination loss:
\begin{align}
    & \min_{W_f} L_D(\mathbf{e}; f),
    \quad \mathbf{e} \in Q, \label{eqn:min-wf} \\
    & L_D(\mathbf{e};f) =
    \begin{cases}
        -\log f(\mathbf{e}), & \mathbf{e} \text{ from source,} \\
        -\log (1-f(\mathbf{e})), & \mathbf{e} \text{ from target,} \\
    \end{cases}
\end{align}
where $L_D$ is a standard classification loss.
In this way, the domain classifier is trained with signals from multiple batches, leading to a faster and more robust estimation of the domain boundary.
% to the optimal domain boundary.
% It is worth noting that since the embeddings stored in Q are detached, gradients will not back propagate to the encoder $g$  but only update $W_f$ as shown in \Cref{eqn:min-wf}.
% In this way, efficiency of the momentum training is greatly improved since redundant forward and backward propagation through the large encoder model are saved.

\subsection{Adversarial Learning for Domain Invariant Representations}
\label{sec:adv}

With an estimated domain boundary from the domain classifier $f$, \model{} adversarially trains the encoder $g$ to generate domain invariant representations that $f$ cannot distinguish, by minimizing an adversarial loss $L_M$.
Here we choose the widely used Confusion loss~\citep{adda}:
% and the adversarial loss for a q--d pair $x$ is:
\begin{equation}
    % l_M(y) &= -\frac{1}{2}\log h(g(y)) -\frac{1}{2}\log (1-h(g(y))), \\
    L_M(x; g, f) = -\frac{1}{2}\Big(\log f(g(q)) +\log (1-f(g(q))) +\log f(g(d)) +\log (1-f(g(d)))\Big),
\end{equation}
where $x \in \{x^s, x^t\}$ is a q-d pair from either source or target, as the confusion loss aims to push for random classification probability for any data points. It reaches the minimum when the embeddings are domain invariant and the domain classifier predict 50\%-50\% probability for all data.

To push for domain invariance, we freeze the domain classifier and update parameters of the encoder:
\begin{equation}
    \min_{\theta_g} \lambda \sum_{x \in \{x^s, x^t\}} L_M(x; g, f) \label{eqn:loss-adversarial}
\end{equation}
We use the hyperparameter $\lambda$ to balance the learning of DR ranking in the source domain (\Cref{eqn:loss-ranking}) and the learning of domain invariance (\Cref{eqn:loss-adversarial}).

To summarize, for each training batch in the source domain, the domain classifier $f$ and the encoder $g$ are optimized by:
\begin{align}
    \min_{W_f}& \quad L_D(\mathbf{e}; f), \quad \mathbf{e} \in Q,\\
    \min_{\theta_g}& \sum_{x^{s+},x^{s-}}
    L_R(r(x^{s+}), r(x^{s-})) + \lambda \sum_{x \in \{x^s, x^t\}} L_M(x; g, f),
\end{align}
where $f$ is trained to estimate the boundary between source/target and $g$ is trained to provide domain invariant representations while capturing the relevance matches in the source domain.

\section{Experimental Setups}
\label{sec:setup}

\header{Datasets}
We choose the MS MARCO passage dataset~\citep{msmarco} as the source domain dataset and choose the 15 publicly available datasets gathered in the BEIR benchmark~\citep{beir} as target domain datasets.
These datasets cover a large number of various domains, including biomedical, finance, scientific, etc.
We treat each target domain dataset separately and produce an individual model for each of them, following standard unsupervised domain adaptation setup~\citep{long2016unsupervised}.
Details of the datasets can be found in \Cref{appx:datasets}.

\header{Evaluation for DR}
Target domain datasets do not always have an ideal coverage for relevance labels.
The annotation procedure of many datasets requires some retrieval models to generate candidates for labeling, which are mainly sparse models at the time of construction.
Therefore, the evaluation of these datasets is not only biased towards sparse models but also less sensitive to dense models.
High Hole rates (a \textit{hole} is a predicted q--d pair without annotation) are often observed for dense models~\citep{ance,beir}.
In fact, ANCE underperforms sparse methods such as BM25
on TREC-COVID with the original annotation, but after adding extra labels based on ANCE's prediction, its scores greatly improve, achieving the state of the art~\citep{beir}.
Nevertheless, TREC-COVID is the dataset with the lowest hole rates for DR models since participating systems include dense ones, and is one of the best to measure the progress of ZeroDR.

\header{Model Validation in ZeroDR}
In the ZeroDR setting, there is no access to relevance labels in the target domain during training/validation.
Therefore, choosing the optimal hyperparameters is impossible without directly tuning on the test set.
In our experiments, most of our hyperparameters are kept the same with ANCE. We also use exactly the same experimental setting and evaluate checkpoints after a fixed number of training steps ($10$k) for all target domain datasets.
This evaluation setup may not yield the optimal empirical results for \model, but it is the closest to ZeroDR in the real world.
Please refer to  \Cref{appx:setting} for detailed hyperparameters.

\header{Baselines}
As a first stage retrieval method, \model's baselines include BM25~\citep{bm25}, DPR~\citep{dpr}, and ANCE~\citep{ance}.
The original DPR is trained on NQ~\citep{nq}, and we train another DPR model on MARCO to eliminate training dataset differences.
The nDCG scores of BM25, DPR-NQ, and ANCE are taken from the BEIR paper (verified to be consistent with our runs); DPR-MARCO and \model{} are from our own evaluation.

BEIR also reports results of other retrieval methods, such as docT5query~\citep{doct5query}, TAS-B~\citep{tas-b}, GenQ~\citep{genq}, ColBERT~\citep{colbert}, etc.
However, they are not directly comparable with \model{} since they may include stronger supervision signals, data augmentation, and/or expensive late interaction, so they are orthogonal with \model{} and can be combined for better empirical results.
Our main baseline is ANCE, which \model{} is built upon and is also shown to be the state of the art on TREC-COVID~\citep{beir}.

\section{Results and Analyses}
\label{sec:results}

This section evaluates the effectiveness of \model{}, its momentum training, and the benefits of domain invariant representations.

\subsection{Effectiveness of Proposed Methods}
\begin{table}[t]
	\centering
	\small
	\caption{
	Overall performance and label coverage (hole rate) on tasks collected in BEIR. Relative improvements of \model{} over its base DR model ANCE is shown in percentages. Datasets are ordered by ANCE's hole rate. Lower hole rates indicate more robust evaluation.
% 	Comparing \model{} against baselines on target domain datasets. \model's improvements are relative to ANCE.
	}
	\label{tab:overall}
% 	\resizebox{\textwidth}{!}{
    \begin{tabular}{@{}l|ccc|c@{\hspace{20pt}}c@{\hspace{0pt}}c@{\hspace{5pt}}cl@{}}
    \hline
	              & \multicolumn{3}{c|}{\textbf{Hole@10}} & \multicolumn{5}{c}{\textbf{nDCG@10}} \\
                   & BM25 & ANCE & \model & BM25 &  
                   \multicolumn{2}{c}{\hspace{-20pt}DPR-(NQ/MARCO)} 
                  & ANCE & \multicolumn{1}{c}{\model} \\
	 \hline
	 TREC-COVID  & 10.6\% & 22.4\% & 19.2\% & 0.616 & 0.332 & 0.561 & 0.654 & 0.676 ($+$3.4\%) \\
	 Touch\'e    & 29.8\% & 56.9\% & 53.5\% & 0.605 & 0.127 & 0.243 & 0.284 & 0.315 ($+$10.9\%) \\
	 DBPedia     & 41.3\% & 65.8\% & 65.0\% & 0.288 & 0.263 & 0.236 & 0.281 & 0.284 ($+$1.1\%) \\
	 NFCorpus    & 74.1\% & 83.1\% & 82.6\% & 0.297 & 0.189 & 0.208 & 0.237 & 0.244 ($+$3.0\%) \\
	 Quora       & 88.7\% & 87.1\% & 87.0\% & 0.742 & 0.248 & 0.842 & 0.852 & 0.856 ($+$0.5\%) \\
	 BioASQ      & 80.7\% & 89.5\% & 89.1\% & 0.514 & 0.127 & 0.232 & 0.306 & 0.320 ($+$4.6\%) \\
	 HotpotQA    & 87.7\% & 90.9\% & 90.7\% & 0.601 & 0.391 & 0.371 & 0.456 & 0.462 ($+$1.3\%) \\
	 FEVER       & 92.6\% & 91.2\% & 91.1\% & 0.648 & 0.562 & 0.589 & 0.669 & 0.680 ($+$1.6\%) \\
	 FiQA        & 93.4\% & 91.5\% & 91.5\% & 0.239 & 0.112 & 0.275 & 0.295 & 0.296 ($+$0.3\%) \\
     ArguAna     & 92.7\% & 92.6\% & 92.6\% & 0.441 & 0.175 & 0.414 & 0.415 & 0.418 ($+$0.7\%) \\
     NQ          & 94.9\% & 92.6\% & 92.6\% & 0.310 & 0.474 & 0.398 & 0.446 & 0.442 ($-$0.9\%) \\
	 SciFact     & 91.5\% & 92.8\% & 92.9\% & 0.620 & 0.318 & 0.478 & 0.507 & 0.502 ($-$1.0\%) \\
	 SCIDOCS     & 92.2\% & 93.8\% & 93.7\% & 0.156 & 0.077 & 0.108 & 0.122 & 0.124 ($+$1.6\%) \\
   Climate-FEVER & 95.7\% & 94.1\% & 93.9\% & 0.179 & 0.148 & 0.176 & 0.198 & 0.206 ($+$4.0\%) \\
	 CQADupStack & 94.8\% & 94.9\% & 94.9\% & 0.316 & 0.153 & 0.281 & 0.296 & 0.297 ($+$0.3\%) \\
    \hline
    \end{tabular}
% 	}
\end{table}

\Cref{tab:overall} shows the overall ZeroDR accuracy of \model{} and baselines on the BEIR benchmark~\citep{beir}.
\model{} improves ANCE's overall effectiveness in the ZeroDR setting.
On datasets with low hole rates (good label coverage), the gains are significant;
on datasets with high hole rates, which are less sensitive to DR model improvement, the gains are less significant but still stable.
Moreover, results of \model{} are obtained \emph{without} hyperparameter tuning or checkpoint selection, and therefore present a fair comparison in the realistic ZeroDR setting.
% Scores from other checkpoints and experimental settings will be shown in \Crefrange{sec:ablation}{sec:emb-space}.

\subsection{Ablation Studies}
\label{sec:ablation}

\begin{table}[t]
	\centering
	\small
	\caption{Ablation studies on TREC-COVID and Touch\'e.
% 	The momentum method is shown to be critical for successful training of domain invariant representation.
    Underlined scores are generated by the default experimental setting, and bold scores are the highest in the row.
    Scores are nDCG@10.
% 	\cx{I don't like using row number to refer to methods. The name of the row should be self-explainable. Also if we transpose this table it is more consistent with Table 1. Group methods by with momentum and with out, also steps}
	}
	\label{tab:ablation}
% 	\resizebox{\columnwidth}{!}{
\begin{tabular}{l@{\qquad}|cc|cc|ccc|c}
\hline
	
 & \multicolumn{2}{c|}{\textbf{w/o Momentum}} & \multicolumn{5}{c|}{\textbf{\model{} Variants w/ Momentum}}  & \multirow{3}{*}{\textbf{ANCE}}
 \\ \cline{0-7}
	
	\textbf{Adversarial Loss}  & \multicolumn{2}{c|}{Confusion} & Minimax & GAN & \multicolumn{3}{c|}{Confusion}  &  \\ \cline{0-7}
	\textbf{Momentum Step $n$}& 1 & 1k  & 1k & 1k & 100 & 1k & 5k \\

% 	$k$     &  &  & 1 & 1000 & 100 & 5000 & 1000 & 1000 \\
	\hline
	TREC-COVID &  	 0.650 & 0.664 & 0.666 & 0.641 & 0.649 & \underline{\bf 0.676} & 0.600 & 0.654 
\\
	Touch\'e    & 0.294 & 0.309 & 0.322 & 0.325 & 0.294 & \underline{0.315} & \bf 0.333 & 0.284
\\
	\hline
	\end{tabular}

% 	\begin{tabular}{l@{\hspace{40pt}}lcc@{\hspace{40pt}}cc}
% 		\toprule
% % 		\# & \multicolumn{3}{c}{Setting} & \multicolumn{2}{c}{Dataset} \\
% % 		\midrule
% 		\# & $L_M$     & Momentum        & $k$  & TREC-COVID & Touch\'e   \\
% 		\midrule
% 		1 & Confusion & \textit{async}  & 1000 & 0.676 & 0.315  \\
% 		2 & Confusion & -               & 1    & 0.650 & 0.294 \\
% 		3 & Confusion & \textit{repeat} & 1000 & 0.664 & 0.309  \\
% 		4 & Confusion & \textit{async}  & 100  & 0.649 & 0.294 \\
% 		5 & Confusion & \textit{async}  & 5000 & 0.600 & 0.333 \\
% 		6 & Minimax   & \textit{async}  & 1000 & 0.666 & 0.322 \\
% 		7 & GAN       & \textit{async}  & 1000 & 0.641 & 0.325 \\
% 		\midrule
% 		\multicolumn{4}{l}{Vanilla ANCE}    & 0.654 & 0.284 \\ \bottomrule
% 	\end{tabular}
% 	}
\end{table}

Our ablation studies evaluate the importance of the momentum method and the effects of other experimental setups. 
We use the two datasets with the best label coverage, TREC-COVID and Touch\'e, and show the results in \Cref{tab:ablation}. \model{}'s default setting is underlined.
% The column with underlined nDCG scores is the default setting.

Firstly, we evaluate the accuracy of \model{} \textit{without} the momentum method, i.e., we do not maintain the momentum queue, but simply update the domain classifier with embeddings of the current batch.
Without momentum, \model's improvement over ANCE diminishes.
% This shows that the momentum method is critical to the successful training of \model.
Secondly, we evaluate \model{} with other two choices of adversarial loss (\Cref{eqn:loss-adversarial}):\ Minimax and GAN~\citep{adda}.
GAN loss is less stable as expected~\citep{adda}, while Minimax performs comparatively to Confusion.
This shows that \model{} can also be applied with other domain adaptation training methods.
Thirdly, we vary the momentum step $n$ without changing the rest experimental settings. We find that $n$ mainly impacts the balance between learning nearest neighbor locality and learning the domain invariance, so it is an important hyperparameter for \model{}.

\subsection{Convergence of Adversarial Training with Momentum}
\label{sec:convergence}
\begin{figure}[t]
\hspace{-5pt}
\begin{subfigure}[]{0.24\textwidth}
     \centering
     \includegraphics[width=\textwidth]{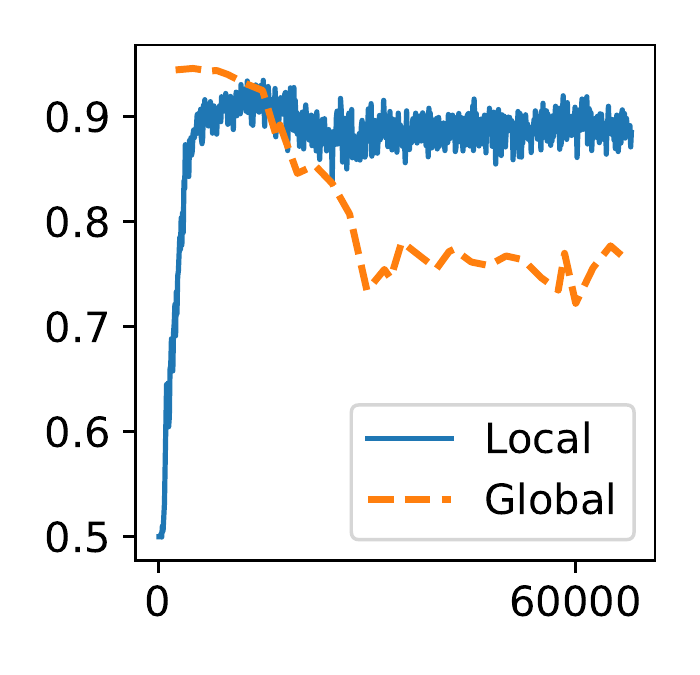}
     \caption{w/ Mom.:\ document}
\end{subfigure}
\begin{subfigure}[]{0.24\textwidth}
     \centering
     \includegraphics[width=\textwidth]{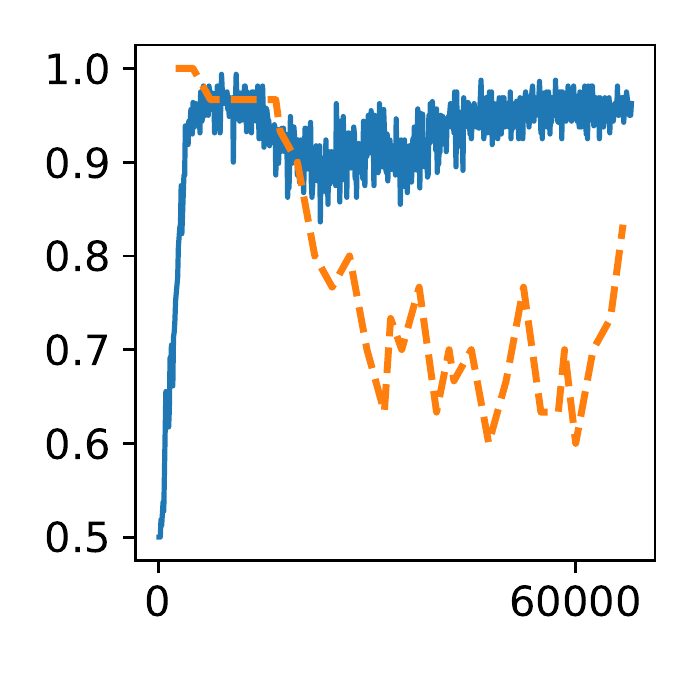}
     \caption{w/ Mom.:\ query}
\end{subfigure}
\begin{subfigure}[]{0.24\textwidth}
     \centering
     \includegraphics[width=\textwidth]{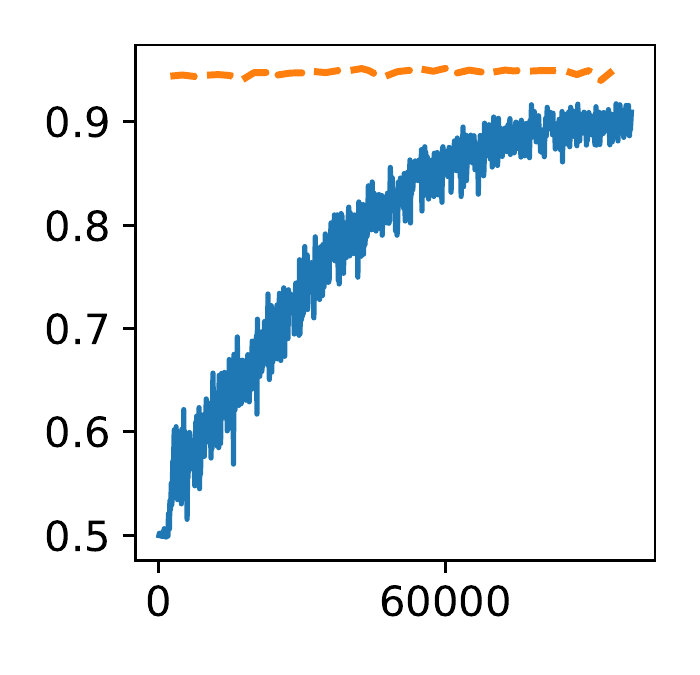}
     \caption{\mbox{w/o Mom.:\ document}}
\end{subfigure}
\begin{subfigure}[]{0.24\textwidth}
     \centering
     \includegraphics[width=\textwidth]{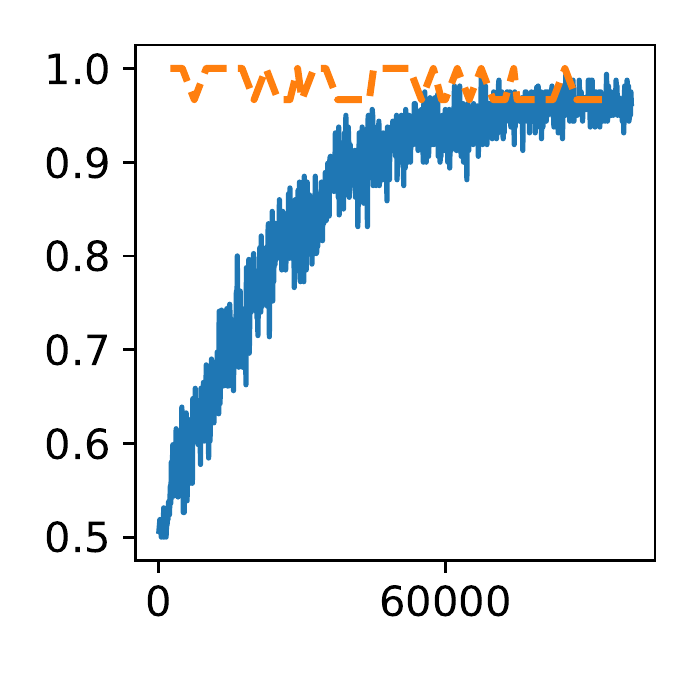}
     \caption{w/o Mom.:\ query}
\end{subfigure}

\caption{Global and Local Domain-Acc with/without momentum (Mom.), at different training steps.}
\label{fig:dry-running}
\end{figure}
In this subsection we evaluate the impact of momentum in adversarial training.
To quantify domain invariance, we use \textit{Domain Classification Accuracy} (Domain-Acc), which includes two measurements based on the choices of domain classifier:
(1) The domain classifier is trained globally on source and target embeddings until convergence, which leads to \textit{Global} Domain-Acc.
(2) We take the domain classifier used in \model's training ($f$ in \Cref{sec:discrimination}), and record its accuracy when it is applied on a new batch, which leads to \textit{Local} Domain-Acc.
Global Domain-Acc measures the real degree of domain invariance:\ it is lower when the embeddings of the two domains are not easily separable.
Local Domain-Acc is an approximation provided by the domain classifier $f$.
A large gap between Local and Global accuracy means that the domain boundary estimated by $f$ is inaccurate.

In \Cref{fig:dry-running}, we compare Global and Local Domain-Acc on the TREC-COVID dataset when momentum is/isn't used .
With momentum, Local Domain-Acc quickly increases to be comparable with Global Domain-Acc.
The domain classifier $f$ (used in \model's training) converges quickly and Global Domain-Acc starts to decrease.
Embeddings from the two domains become less separable as the result of effective adversarial training.
Note that Local Domain-Acc does not decrease because $f$ has seen and memorized almost all data, while Global Domain-Acc's domain classifier is always tested on unseen data.
This shows that momentum helps with the balance of adversarial training, ensuring its convergence towards a domain invariant representation space.

On the other hand, when momentum is not used, there exists a long lasting gap between Local and Global Domain-Acc, showing that $f$ does not capture the domain boundary well.
As a result, the two domains remain (almost) linearly separable in the embedding space, as shown by the fact that Global Domain-Acc does not decrease, and the model fails to learn domain invariance.

\subsection{Impact of Domain Invariant Representation Learning}
\label{sec:emb-space}

\begin{table}[t]
\centering

\caption{K-Nearest Neighbor Source Percentage (KNN-Source\%) and nDCG@10 scores after different number of training steps with/without momentum, on TREC-COVID.}
\label{tab:nsp}
\small
% \resizebox{\textwidth}{!}{
\begin{tabular}{l@{\hspace{30pt}}cccc@{\hspace{20pt}}cccc}
\hline
% \midrule
 & \multicolumn{4}{c}{KNN-Source\%} & \multicolumn{4}{c}{nDCG@10} \\
 \cmidrule(r{15pt}){2-5} \cmidrule(r){6-9}
Checkpoint ($\rightarrow$) & 0 & 10k & 30k & 50k & 0 & 10k & 30k & 50k \\
\hline
% \multicolumn{9}{c}{With Momentum} \\
% \midrule
% nDCG@10 & \multicolumn{2}{c}{\hspace{-35pt}0.654} & \multicolumn{2}{c}{\hspace{-35pt}0.676} & \multicolumn{2}{c}{\hspace{-35pt}0.689} &\multicolumn{2}{c}{\bf \hspace{-15pt}0.724} \\
w/ Momentum  &  5.2\% &  6.2\% & 14.0\% & 17.2\% & 0.654 & 0.676 & 0.689 & \bf 0.724 \\
% Target Queries  &  5.20\% & 94.80\% &  6.16\% & 93.84\% & 13.96\% & 86.04\% & 17.18\% & 82.82\% \\
% Target Passages &  2.69\% & 97.31\% &  2.62\% & 97.38\% &  3.51\% & 96.49\% &  3.16\% & 96.84\% \\
% Source Queries  & 98.94\% &  1.06\% & 99.07\% &  0.93\% & 98.72\% &  1.28\% & 98.60\% &  1.40\% \\
% Source Passages & 95.97\% &  4.03\% & 91.85\% &  8.15\% & 86.05\% & 13.95\% & 86.02\% & 13.98\% \\
% \toprule
% \multicolumn{9}{c}{Without Momentum} \\
% \midrule
% Checkpoint & \multicolumn{2}{c}{\hspace{-35pt}0 Steps} & \multicolumn{2}{c}{\hspace{-35pt}{10k Steps}} & \multicolumn{2}{c}{\hspace{-35pt}{20k Steps}} & \multicolumn{2}{c}{\hspace{-15pt}25k Steps} \\
% nDCG@10 & \multicolumn{2}{c}{\hspace{-35pt}0.654} & \multicolumn{2}{c}{\hspace{-35pt}0.650} & \multicolumn{2}{c}{\hspace{-35pt}0.673} &\multicolumn{2}{c}{\hspace{-15pt}0.668} \\
w/o Momentum  &  5.2\% &  5.4\% & 5.6\% & 5.6\% & 0.654 & 0.650 & 0.673 & 0.668 \\
% Taret Queries  &  5.20\% & 94.80\% &  5.36\% & 94.64\% &  5.56\% & 94.44\% &  5.60\% & 94.40\% \\
% Target Passages &  2.69\% & 97.31\% &  3.60\% & 96.40\% &  3.27\% & 96.73\% &  2.96\% & 97.04\% \\
% Source Queries  & 98.94\% &  1.06\% & 99.24\% &  0.76\% & 99.24\% &  0.76\% & 99.18\% &  0.82\% \\
% Source Passages & 95.97\% &  4.03\% & 95.06\% &  4.94\% & 95.53\% &  4.47\% & 95.30\% &  4.70\% \\
\hline
\end{tabular}
% }

\end{table}

\begin{figure}[t]
\begin{subfigure}[]{0.24\textwidth}
     \centering
     \includegraphics[width=\textwidth]{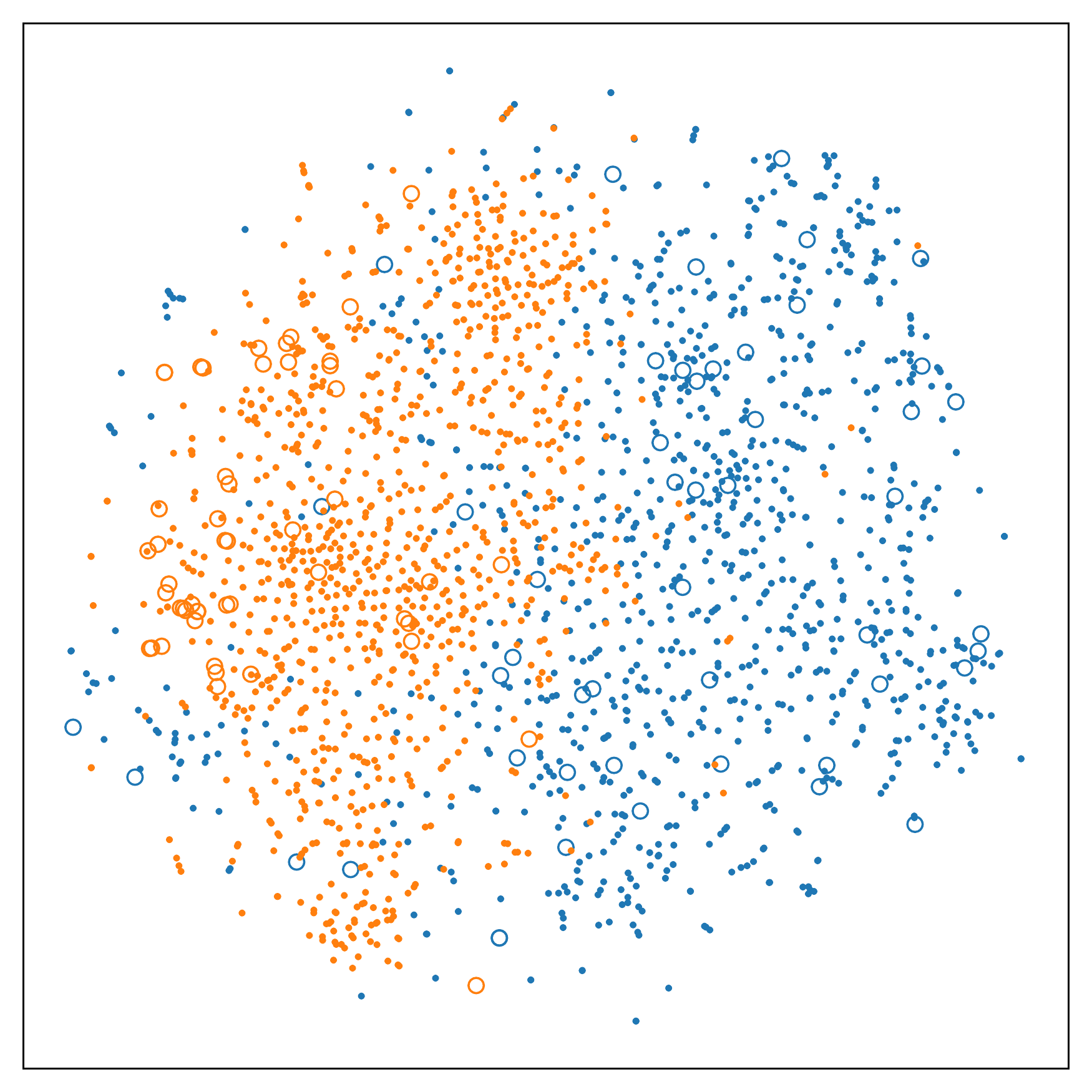}
     \caption{\model{} (0)}
\end{subfigure}
\begin{subfigure}[]{0.24\textwidth}
     \centering
     \hspace{-7.5pt}
     \includegraphics[width=\textwidth]{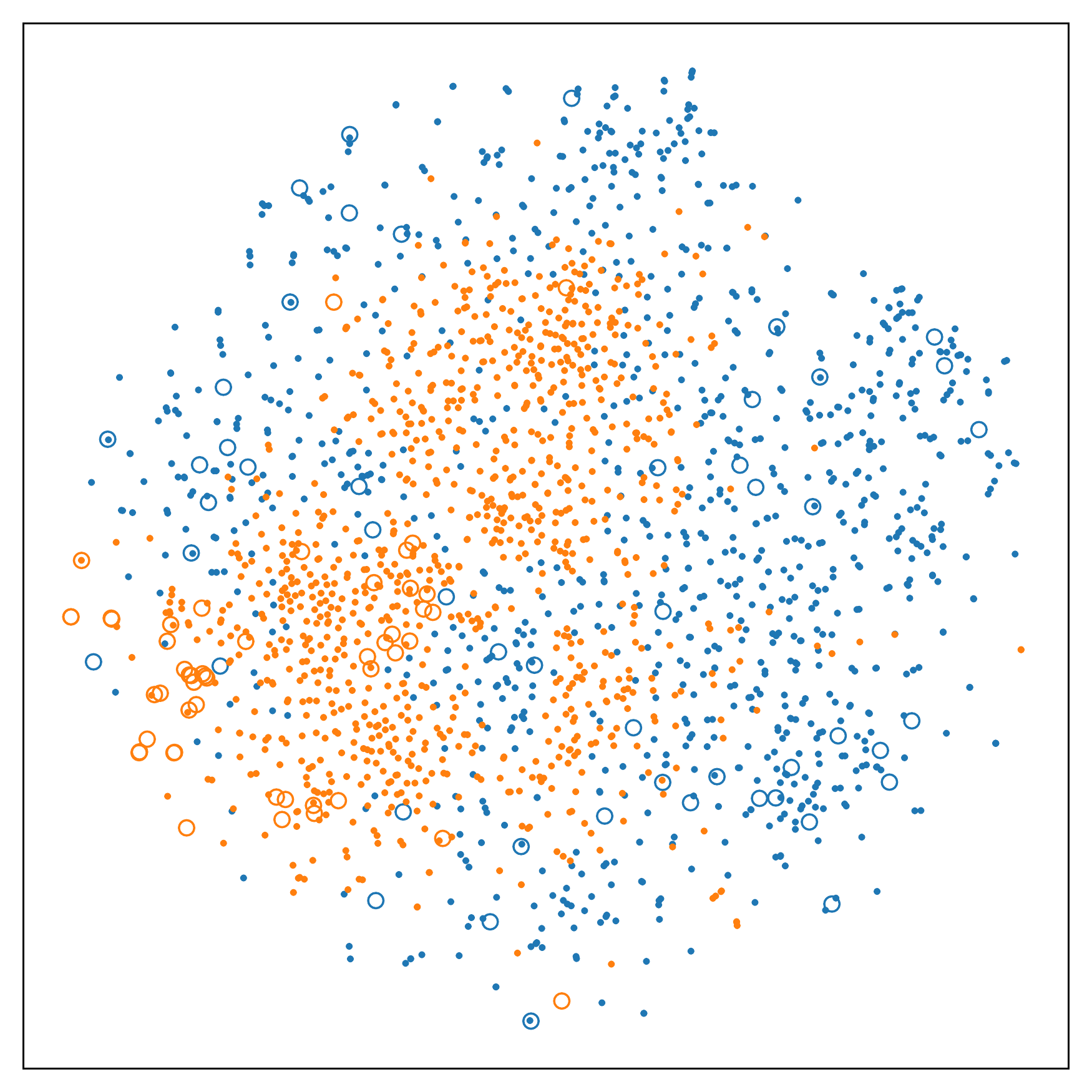}
     \caption{\model{} (10k)}
\end{subfigure}
\begin{subfigure}[]{0.24\textwidth}
     \centering
     \hspace{-15pt}
     \includegraphics[width=\textwidth]{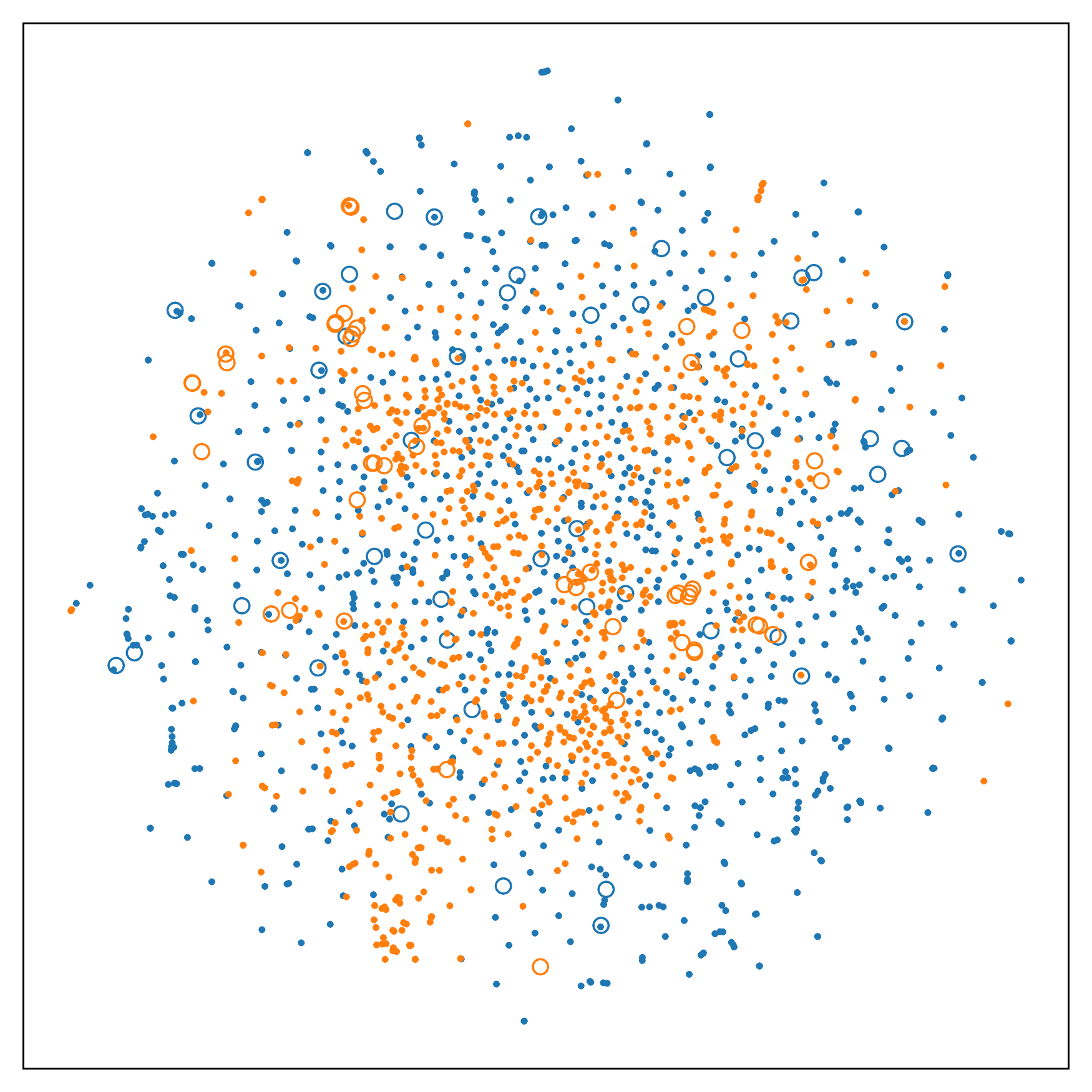}
     \caption{\model{} (30k)}
\end{subfigure}
\begin{subfigure}[]{0.24\textwidth}
     \centering
     \hspace{-20pt}
     \includegraphics[width=\textwidth]{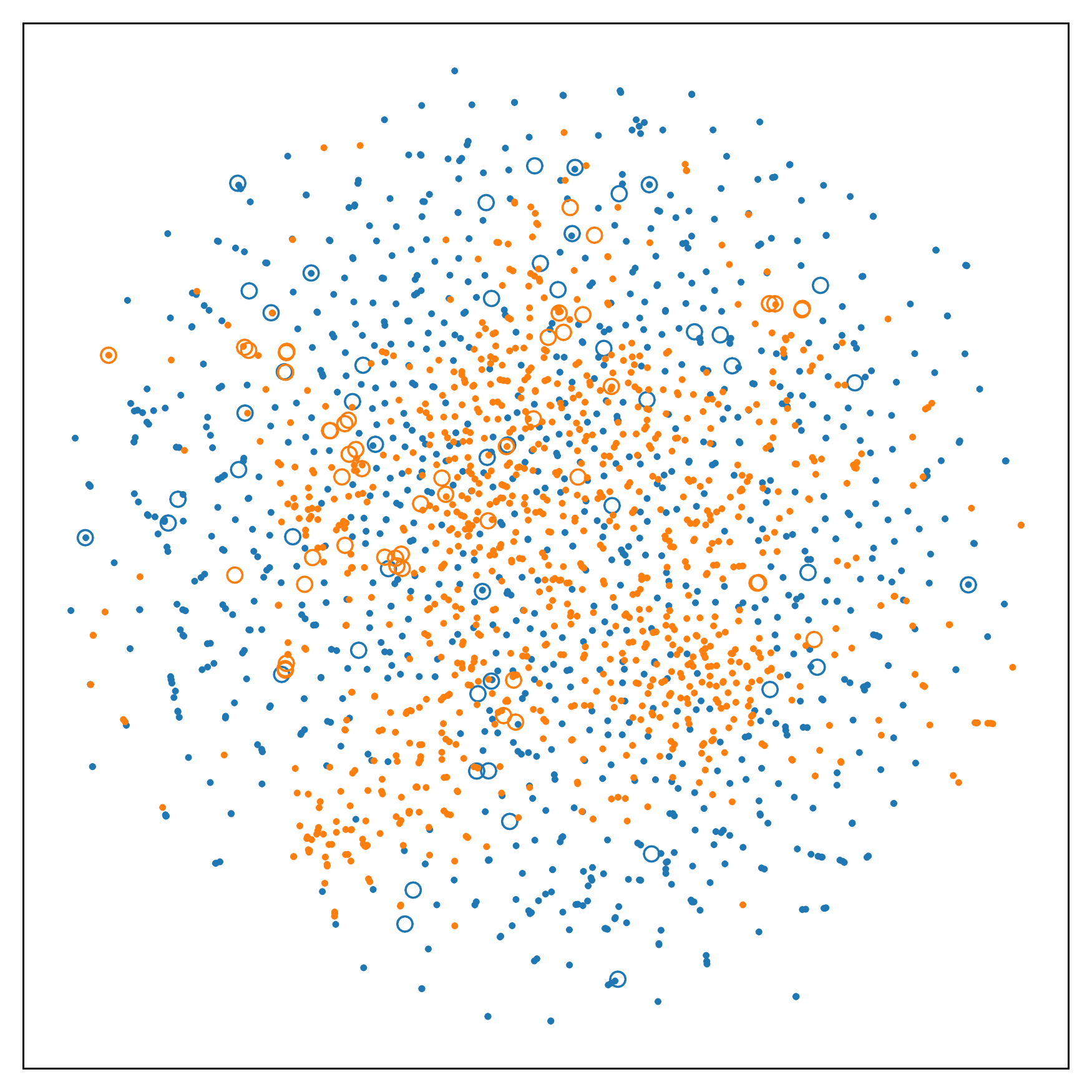}
     \caption{\model{} (50k)}
\end{subfigure}
\begin{subfigure}[]{0.24\textwidth}
     \centering
     \includegraphics[width=\textwidth]{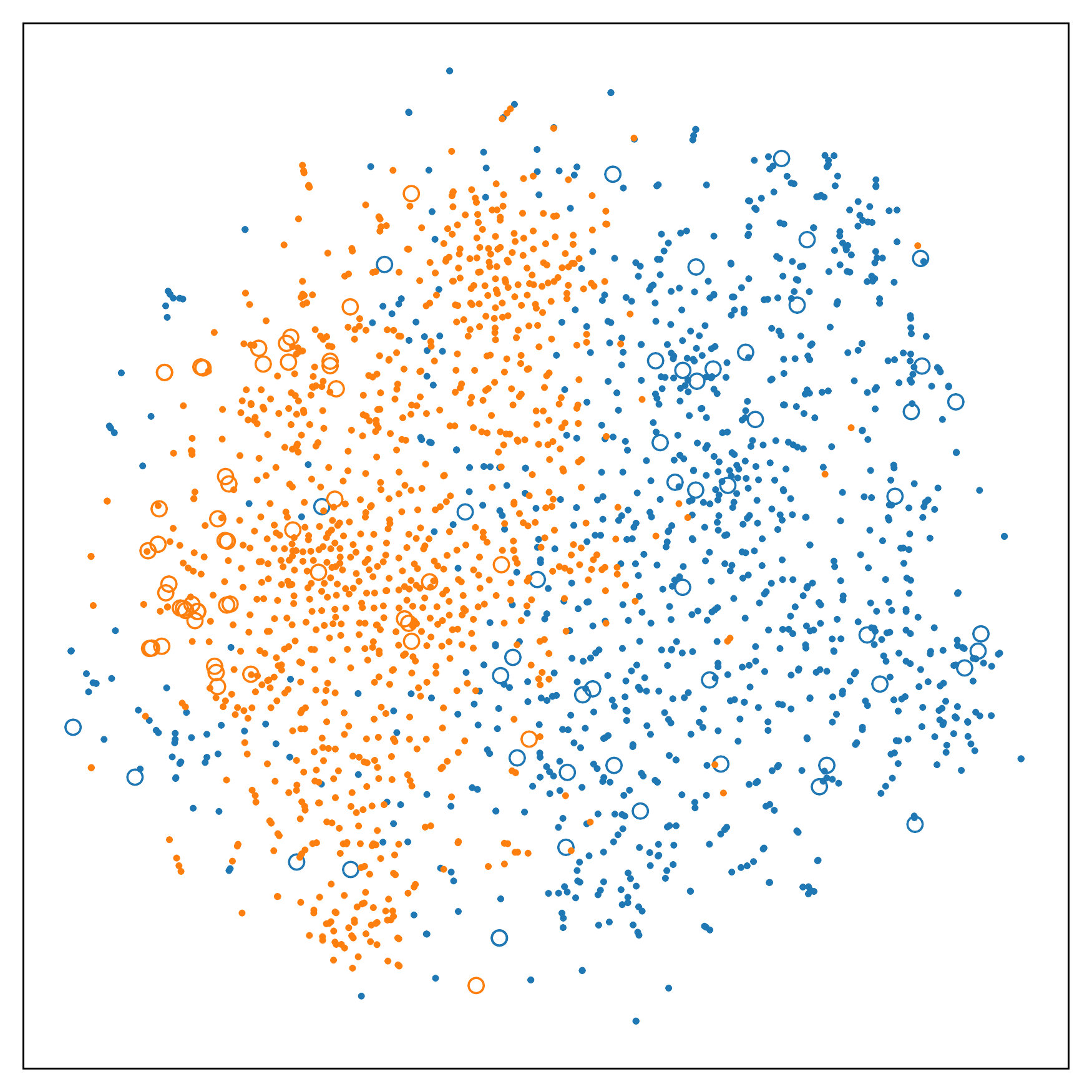}
     \caption{w/o Mom. (0)}
\end{subfigure}
\begin{subfigure}[]{0.24\textwidth}
     \centering
     \includegraphics[width=\textwidth]{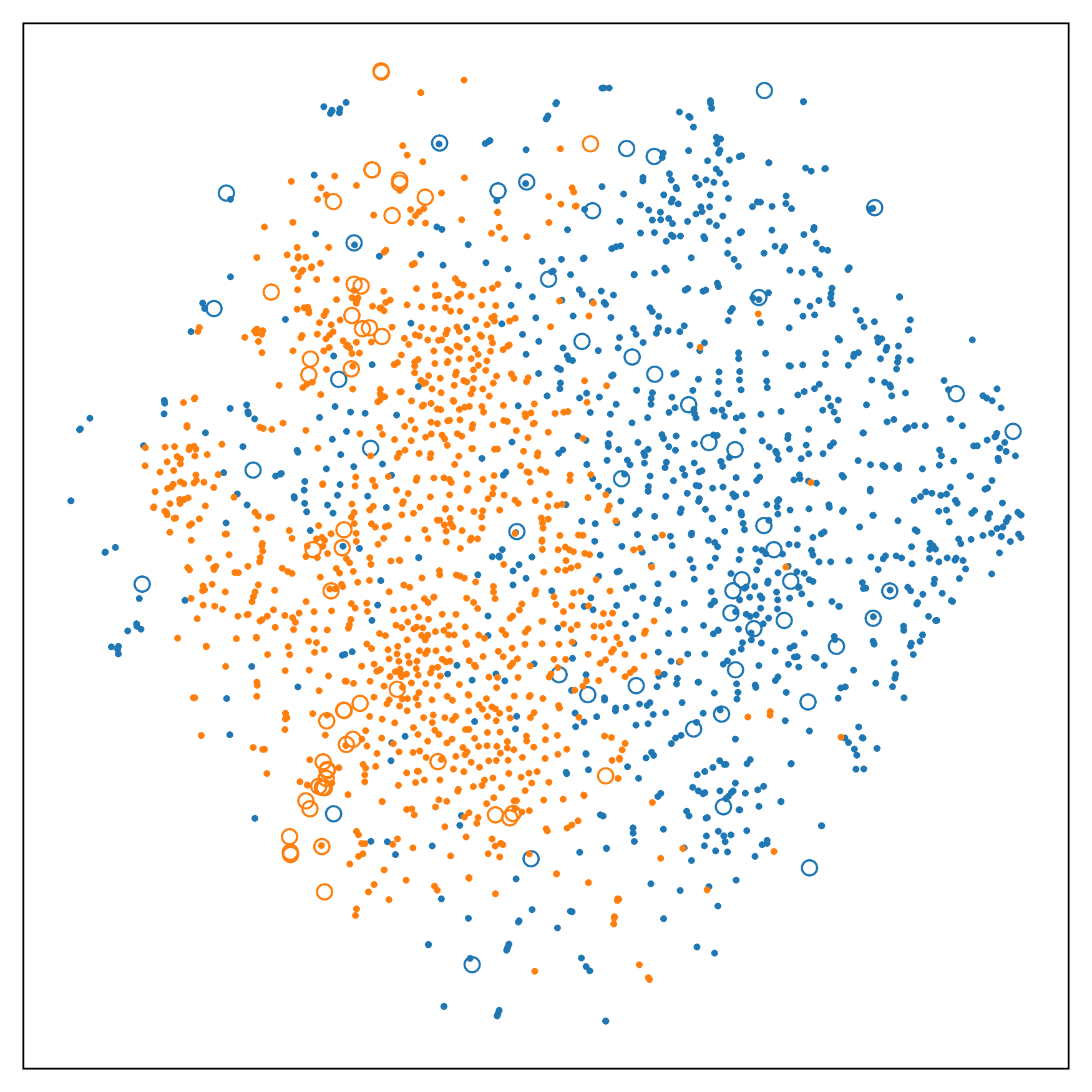}
     \caption{w/o Mom. (10k)}
\end{subfigure}
\begin{subfigure}[]{0.24\textwidth}
     \centering
     \includegraphics[width=\textwidth]{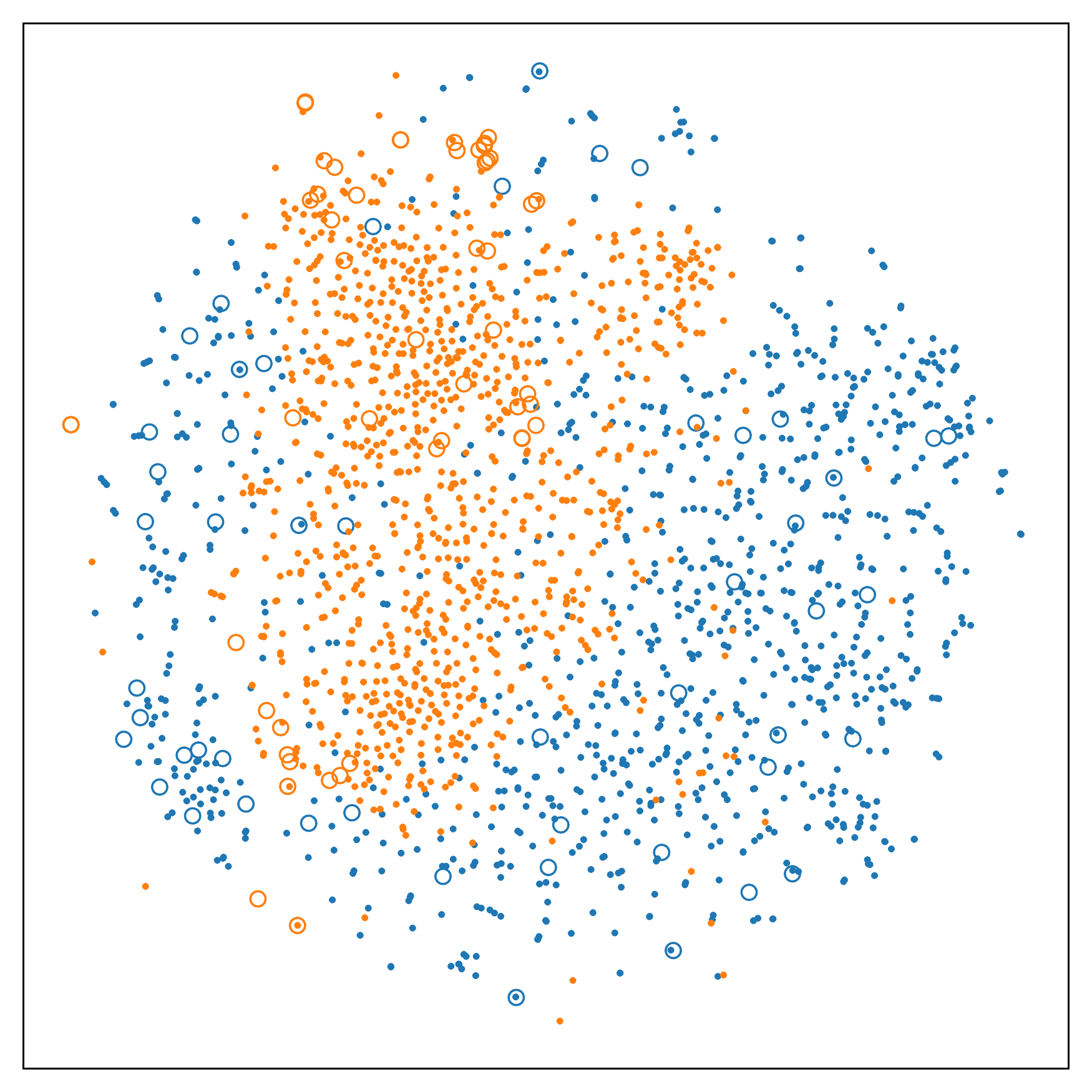}
     \caption{w/o Mom. (30k)}
\end{subfigure}
\begin{subfigure}[]{0.24\textwidth}
     \centering
     \includegraphics[width=\textwidth]{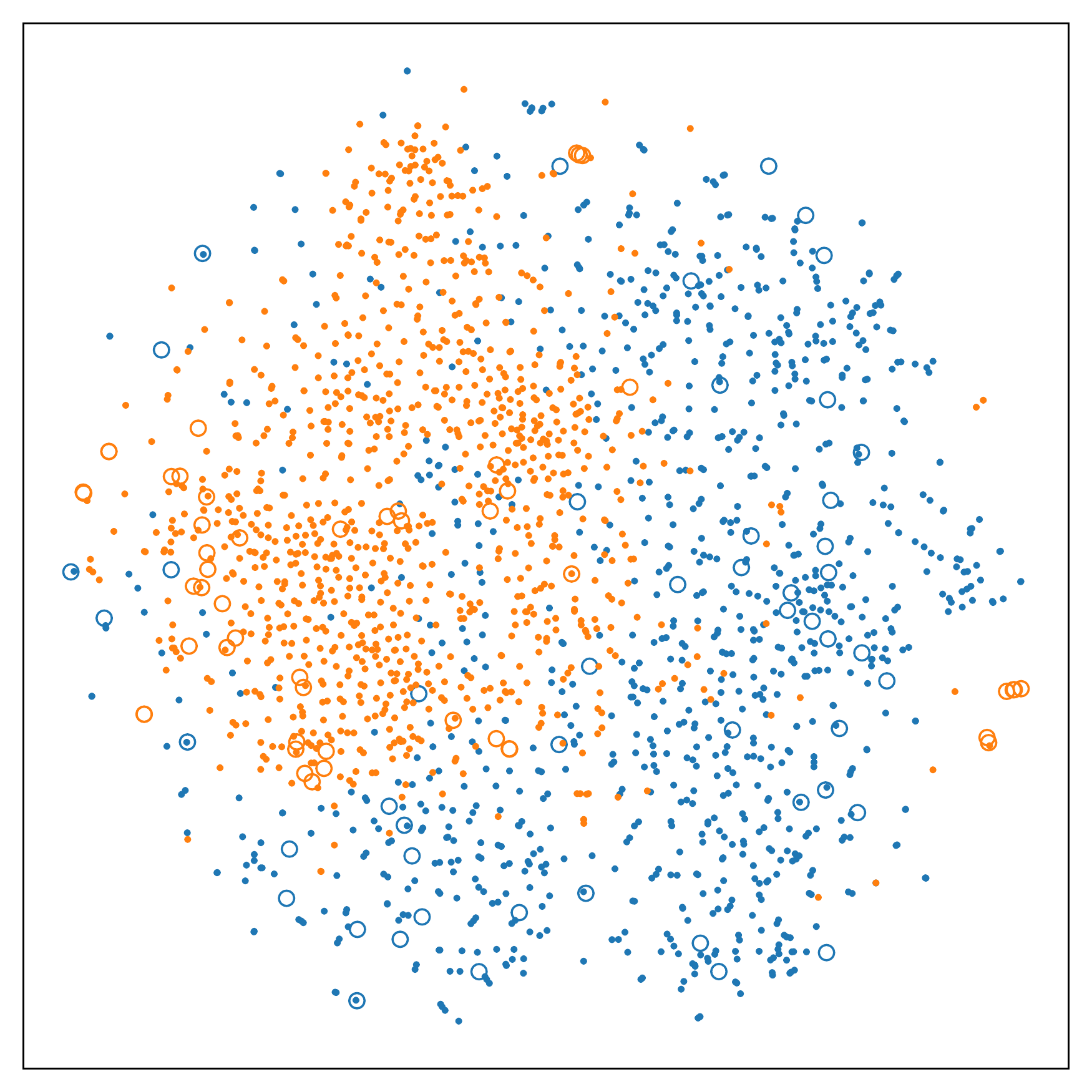}
     \caption{w/o Mom. (50k)}
\end{subfigure}
\caption{T-SNE of the representation space after different training steps (in the parentheses), with/without momentum. Blue:\ source (MARCO); orange:\ target (TREC-COVID).
}
\label{fig:tsne}
\end{figure}

\begin{figure}[t]
\begin{subfigure}[]{0.24\textwidth}
     \centering
     \includegraphics[width=1\textwidth]{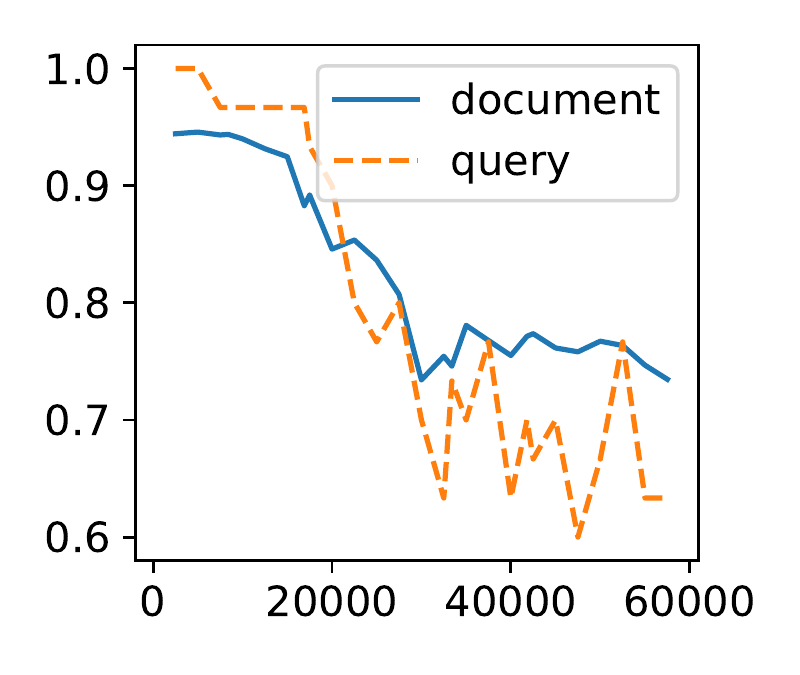}
     \caption{TREC-COVID\\\centering Global Domain-Acc}
\end{subfigure}
\begin{subfigure}[]{0.24\textwidth}
     \centering
     \includegraphics[width=1\textwidth]{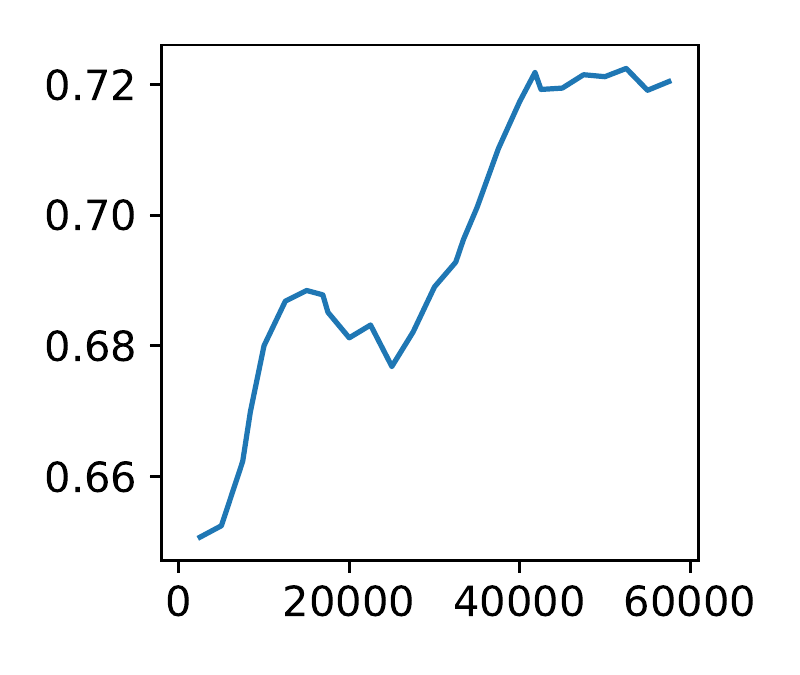}
     \caption{TREC-COVID\\\centering nDCG@10}
\end{subfigure}
\begin{subfigure}[]{0.24\textwidth}
     \centering
     \includegraphics[width=1\textwidth]{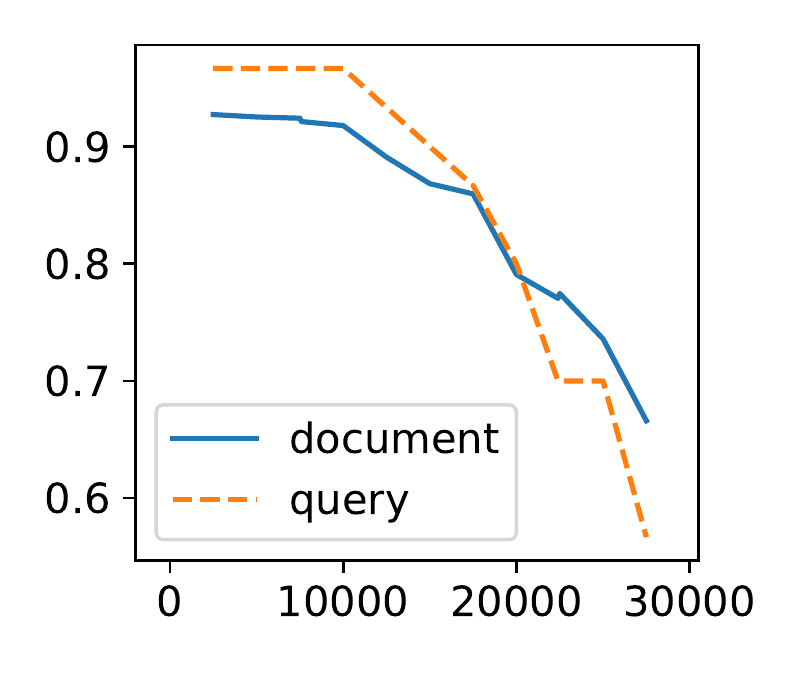}
     \caption{Touch\'e\\\centering Global Domain-Acc}
\end{subfigure}
\begin{subfigure}[]{0.24\textwidth}
     \centering
     \includegraphics[width=1\textwidth]{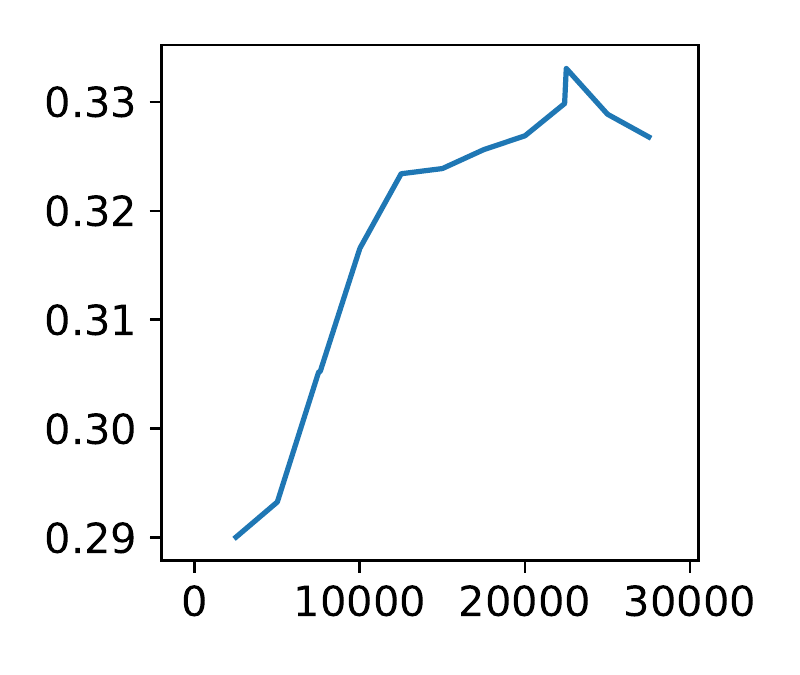}
     \caption{Touch\'e\\  nDCG@10}
\end{subfigure}

\caption{Global Domain-Acc and target domain ZeroDR accuracy at different training steps.
% \cx{the y-axes labels are hard to match to figures. i got confused and thought marco mrr is for (b)} why DCA not Global-Domain-Acc
}
\label{fig:ndcg-dry}
\end{figure}

In this subsection we study the behavior and benefits of \model{} in learning domain invariance. We focus on TREC-COVID as it provides the most robust evaluation for ZeroDR.

\header{Learning Domain Invariance with Momentum}
We show how the momentum method gradually pushes for a domain invariant representation space.
To measure how much the two domains are mixed together, we use \textit{K-Nearest Neighbor Source Percentage} (\textit{KNN-Source\%}):
We index source and target documents together;
given a target domain query in the embedding space, we retrieve its top-100 nearest documents from the index, and calculate the percentage of source documents from the nearest neighbors;
the average percentage for all target domain queries is reported.
A higher KNN-Source\% indicates that the target domain embeddings are more mixed with source domain ones, indicating a more domain invariant representation space.

The results are shown in \Cref{tab:nsp}.
With momentum, both KNN-Source\% and nDCG gradually increase as training proceeds.
This shows that when target domain embeddings are pushed towards the source domain, ranking performance of the target domain also improves.
On TREC-COVID, \model{} eventually reaches a state-of-the-art \textbf{0.724} for first stage retrievers.
On the other hand, without momentum, KNN-Source\% and nDCG scores hardly increase.

We also use t-SNE~\citep{tsne} to visualize the learned representation space at different training steps in \Cref{fig:tsne}.
Before training with \model{}, the two domains are well separated in the representation space learned by ANCE.
With more \model{} training steps, the target domains are pushed towards the source domain and gradually becomes a subset of it.
Without momentum, the two domains remain separated, as observed in \Cref{tab:nsp}.

\begin{table}[t]
\centering
\caption{Case study:\ nearest source queries of a target query before and after \model{} training.}
\label{tab:case}
% \small
\resizebox{\textwidth}{!}{%
\begin{tabular}{l|l|r}
\hline
Target & what are the transmission routes of coronavirus? & nDCG@10 gain: 0.23
\\ \hline
\makecell[l]{Source\\Before} &
\multicolumn{2}{c}{\makecell[l]{
$\bullet$ what is the coronavirus \hspace{90pt} $\bullet$ incubation period for coronavirus \\
$\bullet$ what are symptoms of coronavirus
}} \\ \hline
\makecell[l]{Source\\After} &
\multicolumn{2}{c}{\makecell[l]{
$\bullet$ countries where guinea worm is transmitted
\quad $\bullet$ what is the most common method of hiv transmission \\
$\bullet$ through which body system are cancer cells able to travel to different locations in the body?
}} \\ 
\hline\hline
Target & what is known about an mRNA vaccine for the SARS-CoV-2 virus? & nDCG@10 gain: $-$0.12
\\ \hline
\makecell[l]{Source\\Before} &
\multicolumn{2}{c}{\makecell[l]{
$\bullet$ is there a vaccine for hepatitis
\hspace{63pt} $\bullet$ is there a vaccine for tuberculosis \\
$\bullet$ shingles vaccination needed for those without chickenpox
}} \\ \hline
\makecell[l]{Source\\After} &
\multicolumn{2}{c}{\makecell[l]{
$\bullet$ what makes rna
\hspace{118pt} $\bullet$ what is used to make mrna \\
$\bullet$ what is the mmr vaccine called
}} \\ \hline
\end{tabular}%
}
\end{table}

\header{ZeroDR Accuracy Versus Domain Invariance}
We study the correlation between ZeroDR accuracy and domain invariance. We use Global Domain-Acc as the indicator of domain invariance and plot it with the corresponding ZeroDR accuracy during training in  \Cref{fig:ndcg-dry}.

Global Domain-Acc starts at near 100\%, showing that source and target embeddings are linearly separable with the one-layer domain classifier.
It decreases as training proceeds, and when the learned representation space becomes more domain invariant, the ZeroDR accuracy in the target domain improves alongside.
This shows that domain invariance is the source of improvements of ZeroDR's effectiveness.
We also record that the DR accuracy on the source domain (MARCO) decreases by no more than $0.5\%$.
This indicates that the high dimensional embedding space has sufficient capacity to learn domain invariant representations while maintaining relevance matching in the source domain.

\subsection{Case Study}

We show two examples of queries from TREC-COVID and their nearest MARCO queries before and after \model{} training in \Cref{tab:case}.
In the first case, \model{} pays more attention to ``transmission'', and potentially retrieves more documents about transmission of diseases, thereby improving the nDCG score; documents about ``coronavirus'' also are likely to be retrieved by \model{} since it is a very noticeable word.
In the second case, it focuses on ``mRNA'' more than ``vaccine''.
However, since the mRNA vaccine is relatively new\footnote{In December 2020, the Pfizer–BioNTech COVID vaccine became the first approved mRNA vaccine, according to \url{https://en.wikipedia.org/wiki/MRNA_vaccine}.}
with few appearances in the MARCO dataset, the shift in focus fails to improve \model's effectiveness for this query.

These examples help reveal the source of generalization ability on ZeroDR.
For the DR models to be able to generalize, the source domain itself needs to include information that covers the relevance needs of the target domain;
if there is no such information, as in the second example, generalization becomes a challenge.
Where the source domain has such coverage, \model{} is able to align target queries to source ones with similar information needs in its domain invariant representation space, and such alignments enable DR models to generalize.

\section{Related Work}
\label{sec:related}

In this section, we recap related work in dense retrieval and adversarial domain adaptation.

\header{Dense Retrieval}
Compared to conventional sparse methods for first stage retrieval, dense retrieval (DR) with Transformer-based models~\citep{transformer} such as BERT~\citep{bert} and RoBERTa~\citep{roberta} conduct retrieval in the dense embedding space~\citep{ict,Chang2020Pre-training,guu2020realm,dpr,luan2021sparse}.
Compared with its sparse counterparts, DR improves retrieval efficiency and also provides comparable or even superior effectiveness for in-domain datasets.

One of the most important research questions for DR is how to obtain meaningful negative training instances.
DPR~\citep{dpr} uses BM25 to find stronger negatives in addition to in-batch random negatives.
RocketQA~\citep{rocketqa} uses cross-batch negatives and also filters them with a strong reranking model.
ANCE~\citep{ance} uses an asynchronously updated negative index of the being-trained DR model to retrieve global hard negatives.
% and has achieved the state of the art.

Recently, the challenges of DR models' generalization in the zero-shot setting has attracted much attention~\citep{beir,mrtydi,li2021encoder}.
One way to improve ZeroDR is by synthetic query generation~\citep{liang2020embedding,genq}, which first trains a doc2query model that learns to generate queries in the source domain given their relevant documents, and then applies the NLG model on target domain documents to generate queries.
The target domain documents and generated queries form weak supervision labels in the target domain to train DR models.
Our method differs from these approaches and focuses on directly improving the generalization ability of the learned representation space.

\header{Adversarial Domain Adaptation}
Unsupervised domain adaptation (UDA) has been studied extensively for computer vision applications.
For example, maximum mean discrepancy~\citep{Long_2013_ICCV,tzeng2014deep,sun2016deep} measures domain difference with a pre-defined metric and explicitly minimizes the difference.
Following the advent of GAN~\citep{gan}, adversarial training for UDA is proposed:\ an auxiliary domain classifier learns to discriminate source and target domains, while the main classifier model is adversarially trained to confuse the domain classifier~\citep{dann,bousmalis2016domain,adda,luo2017label}.
The adversarial method does not require pre-defining the domain difference metric, allowing more flexible domain adaptation.
\model{} builds upon the success of these UDA methods and introduces a new momentum learning technique that is necessary to learn domain invariant representations in the ZeroDR setting.

\section{Conclusion and Future Work}
\label{sec:conclusion}

In this paper, we present \model{}, a new representation learning method that improves the zero-shot generalization ability of dense retrieval models.
We first show that dense retrieval models differ from classification models in their emphases of locality in the representation space.
Then we present a momentum-based adversarial training method that robustly pushes text encoders to provide a more domain invariant representation space for dense retrieval.
Our experiments on ranking datasets from the BEIR benchmark demonstrate robust and significant improvements of \model{} on the zero-shot accuracy of ANCE, a recent state-of-the-art DR model. 

We conduct a series of studies to show the effects of our momentum method in learning domain invariant representations.
Without momentum, the adversarial learning is unstable as the inherent variance of the DR embedding space hinders the convergence of the domain classifier.
With momentum training, the model is able to fuse the target domain data into the source domain representation space, and thus discovers related information from the source domain and improves generalization, without requiring any target domain training labels.

We view \model{} as an initial step of zero-shot dense retrieval, an area demanding democratization of the rapid advancements to many real-world scenarios.
Our approach inherits the success of domain adaptation techniques and upgrades them by addressing the unique challenges of ZeroDR. 
How to better understand the dynamics of representation learning for DR and further improve its effectiveness, robustness, and generalization ability is a future research direction with potential impacts in both representation learning research and also real-world applications.

\section{Reproducibility Statement}

We provide the following information to ensure our proposed method is reproducible:
\begin{itemize}
    \item All datasets are publicly available and details can be found in \Cref{sec:setup} and \Cref{appx:datasets}.
    \item Detailed experimental setups can be found in \Cref{appx:setting}.
    \item Model validation and evaluation details are discussed in \Cref{sec:setup}.
    \item Source code and model checkpoints will be made public when the paper is published.
\end{itemize}

\bibliography{main}
\bibliographystyle{iclr22}

\newpage
\appendix

\section{Datasets Details}
\label{appx:datasets}
Target domain datasets used in our experiments are from the following domains:
\begin{itemize}
    \item General-domain (Wikipedia): DBPedia~\citep{dbpedia}, HotpotQA~\citep{hotpotqa}, FEVER~\citep{fever}, and NQ~\citep{nq}.
    \item Bio-medical: TREC-COVID~\citep{treccovid}, NFCorpus~\citep{nfcorpus}, and BioASQ~\citep{bioasq}.
    \item Finance: FiQA~\citep{fiqa}.
    \item Controversial arguments: Touch\'e~\citep{touche} and ArguAna~\citep{arguana}.
    \item Duplicate questions: Quora~\citep{beir} and CQADupStack~\citep{cqadupstack}.
    \item Scientific: SciFact~\citep{scifact}, SCIDOCS~\citep{scidocs}, and Climate-FEVER~\citep{climatefever}
\end{itemize}

\section{Detailed Experimental Settings}
\label{appx:setting}
We follow the design of ANCE for DR encoder's modeling and training.
We initialize the encoder with the publicly released ANCE checkpoint\footnote{\url{https://github.com/microsoft/ANCE}.}, and randomly initialize the domain classifier.
Detailed hyperparameter choices are shown in \Cref{tab:hp}.
We also use an exponential decay routine for the hyperparameter $\lambda$ to improve training stability, where the value is reduced continuously and shrunk to a half every 10k steps.

\begin{table}[h]
    \centering
	\caption{
    Detailed hyperparameter choices of \model.
    }
	\label{tab:hp}
% 	\resizebox{\textwidth}{!}{
    \begin{tabular}{l@{\hspace{50pt}}c}
    \hline
    Hyperparameter & Value \\
    \hline 
    \multicolumn{2}{c}{Same as ANCE} \\
    \hline
    Learning rate for $\theta_g$ & 1e-6 \\
    Effective batch size & 16 \\
    Maximum Query Length & 64 \\
    Maximum Document Length & 512 \\
    \hline
    \multicolumn{2}{c}{New for \model} \\
    \hline
    Learning rate for $W_f$ & 5e-6 \\
    Early stopping steps & 10k \\
    Momentum step $N$ & 1k \\
    Initial $\lambda$ & 1.0 \\
    \hline
    \end{tabular}
% 	}
\end{table}

\end{document}